\documentclass[iop]{emulateapj}

\usepackage{times} 
\usepackage{graphicx} 
\usepackage{times} 
\usepackage{natbib} 
\usepackage{rotating}

\newcommand {\apgt} {\ {\raise-.5ex\hbox{$\buildrel>\over\sim$}}\ }
\newcommand {\aplt} {\ {\raise-.5ex\hbox{$\buildrel<\over\sim$}}\ }

\makeatother \lefthead{{\sc Tracing the mass lensing the CMB with quasars}}
\righthead{\sc Geach et al.}

\def\Hert{1}
\def\McGill{2}
\def\Dartmouth{3}
\def\KICPChicago{4}
\def\PhysicsUChicago{5}
\def\ArgonneHEP{6}
\def\Miss{7}
\def\UChicago{8}
\def\EFIChicago{9}
\def\AAUChicago{10}
\def\NIST{11}
\def\Berkeley{12}
\def\ColoradoH{13}
\def\Davis{14}
\def\LBNL{15}
\def\Caltech{16}
\def\Arizona{17}
\def\Michigan{18}
\def\Munich{19}
\def\ExcellenceCluster{20}
\def\MPE{21}
\def\CaseWestern{22}
\def\Wyoming{23}
\def\Minnesota{24}
\def\ArtInstChicago{25}
\def\CfA{26}
\def\DI{27}
\def\Toronto{28}
\def\BCCP{28}

\begin{document}

\title{A direct measurement of the linear bias of mid-infrared-selected quasars at z$\approx$1 using Cosmic Microwave Background lensing}

\author{
J.~E.~Geach\altaffilmark{\Hert, \McGill},
R.~C.~Hickox\altaffilmark{\Dartmouth},
L.~E.~Bleem \altaffilmark{\KICPChicago,\PhysicsUChicago,\ArgonneHEP}, 
M.~Brodwin\altaffilmark{\Miss},
G.~P.~Holder\altaffilmark{\McGill},
K.~A.~Aird\altaffilmark{\UChicago},
B.~A.~Benson\altaffilmark{\KICPChicago,\EFIChicago},
S.~Bhattacharya\altaffilmark{\KICPChicago,\ArgonneHEP},
J.~E.~Carlstrom\altaffilmark{\KICPChicago,\EFIChicago,\PhysicsUChicago,\AAUChicago,\ArgonneHEP},
C.~L.~Chang\altaffilmark{\KICPChicago,\EFIChicago,\ArgonneHEP},
H-M.~Cho\altaffilmark{\NIST},
T.~M.~Crawford\altaffilmark{\KICPChicago,\AAUChicago},
A.~T.~Crites\altaffilmark{\KICPChicago,\AAUChicago},
T.~de~Haan\altaffilmark{\McGill},
M.~A.~Dobbs\altaffilmark{\McGill},
J.~Dudley\altaffilmark{\McGill},
E.~M.~George\altaffilmark{\Berkeley},
K.~N.~Hainline\altaffilmark{\Dartmouth},
N.~W.~Halverson\altaffilmark{\ColoradoH},
W.~L.~Holzapfel\altaffilmark{\Berkeley},
S.~Hoover\altaffilmark{\KICPChicago,\PhysicsUChicago},
Z.~Hou\altaffilmark{\Davis},
J.~D.~Hrubes\altaffilmark{\UChicago},
R.~Keisler\altaffilmark{\KICPChicago,\PhysicsUChicago},
L.~Knox\altaffilmark{\Davis},
A.~T.~Lee\altaffilmark{\Berkeley,\LBNL},
E.~M.~Leitch\altaffilmark{\KICPChicago,\AAUChicago},
M.~Lueker\altaffilmark{\Caltech},
D.~Luong-Van\altaffilmark{\UChicago},
D.P.~Marrone\altaffilmark{\Arizona},
J.~J.~McMahon\altaffilmark{\Michigan},
J.~Mehl\altaffilmark{\KICPChicago,\ArgonneHEP},
S.~S.~Meyer\altaffilmark{\KICPChicago,\EFIChicago,\PhysicsUChicago,\AAUChicago},
M.~Millea\altaffilmark{\Davis},
J.~J.~Mohr\altaffilmark{\Munich,\ExcellenceCluster,\MPE},
T.~E.~Montroy\altaffilmark{\CaseWestern},
A.~D.~Myers\altaffilmark{\Wyoming}
S.~Padin\altaffilmark{\KICPChicago,\AAUChicago,\Caltech},
T.~Plagge\altaffilmark{\KICPChicago,\AAUChicago},
C.~Pryke\altaffilmark{\Minnesota},
C.~L.~Reichardt\altaffilmark{\Berkeley},
J.~E.~Ruhl\altaffilmark{\CaseWestern},
J.~T.~Sayre\altaffilmark{\CaseWestern},
K.~K.~Schaffer\altaffilmark{\KICPChicago,\EFIChicago,\ArtInstChicago},
L.~Shaw\altaffilmark{\McGill},
E.~Shirokoff\altaffilmark{\Berkeley},
H.~G.~Spieler\altaffilmark{\LBNL},
Z.~Staniszewski\altaffilmark{\CaseWestern},
A.~A.~Stark\altaffilmark{\CfA},
K.~T.~Story\altaffilmark{\KICPChicago,\PhysicsUChicago},
A.~van~Engelen\altaffilmark{\McGill},
K.~Vanderlinde\altaffilmark{\McGill,\DI,\Toronto},
J.~D.~Vieira\altaffilmark{\Caltech}, 
R.~Williamson\altaffilmark{\KICPChicago,\AAUChicago}, and
O.~Zahn\altaffilmark{\BCCP}
}

{

\altaffiltext{\Hert}{Centre for Astrophysics Research, Science \& Technology Research
Institute, University of Hertfordshire, Hatfield, AL10 9AB, UK j.geach@herts.ac.uk}
\altaffiltext{\McGill}{Department of Physics, McGill University, Montreal, Quebec H3A 2T8, Canada}
\altaffiltext{\Dartmouth}{Department of Physics and Astronomy, Dartmouth College,
6127 Wilder Laboratory, Hanover, NH, 03755, USA}
\altaffiltext{\KICPChicago}{Kavli Institute for Cosmological Physics, University of Chicago, Chicago, IL, 60637, USA}
\altaffiltext{\PhysicsUChicago}{Department of Physics, University of Chicago, Chicago, IL, 60637, USA}
\altaffiltext{\ArgonneHEP}{High Energy Physics Division, Argonne National Laboratory, Argonne, IL, 60440, USA}
\altaffiltext{\Miss}{Department of Physics and Astronomy, University of Missouri, Kansas City, MO, 64110, USA}
\altaffiltext{\UChicago}{University of Chicago, Chicago, IL, 60637, USA}
\altaffiltext{\EFIChicago}{Enrico Fermi Institute, University of Chicago, Chicago, IL, 60637, USA}
\altaffiltext{\AAUChicago}{Department of Astronomy and Astrophysics, University of Chicago, Chicago, IL, 60637, USA}
\altaffiltext{\NIST}{NIST Quantum Devices Group, Boulder, CO, 80305, USA}
\altaffiltext{\Berkeley}{Department of Physics, University of California, Berkeley, CA, 94720, USA}
\altaffiltext{\ColoradoH}{Department of Astrophysical and Planetary Sciences and Department of Physics, University of Colorado, Boulder, CO, 80309, USA}
\altaffiltext{\Davis}{Department of Physics, University of California, Davis, CA, 95616, USA}
\altaffiltext{\LBNL}{Physics Division, Lawrence Berkeley National Laboratory, Berkeley, CA, 94720, USA}
\altaffiltext{\Caltech}{California Institute of Technology, Pasadena, CA, 91125, USA}
\altaffiltext{\Arizona}{Steward Observatory, University of Arizona, Tucson, AZ, 85721, USA}
\altaffiltext{\Michigan}{Department of Physics, University of Michigan, Ann  Arbor, MI, 48109, USA}
\altaffiltext{\Munich}{Department of Physics, Ludwig-Maximilians-Universit\"{a}t, 81679 M\"{u}nchen, Germany}
\altaffiltext{\ExcellenceCluster}{Excellence Cluster Universe, 85748 Garching, Germany}
\altaffiltext{\MPE}{Max-Planck-Institut f\"{u}r extraterrestrische Physik, 85748 Garching, Germany}
\altaffiltext{\CaseWestern}{Physics Department, Center for Education and Research in Cosmology and Astrophysics, Case Western Reserve University, Cleveland, OH, 44106, USA}
\altaffiltext{\Wyoming}{Department of Physics and Astronomy, University of Wyoming, Laramie, WY, 82072, USA}
\altaffiltext{\Minnesota}{Department of Physics, University of Minnesota, Minneapolis, MN, 55455, USA}
\altaffiltext{\ArtInstChicago}{Liberal Arts Department, School of the Art Institute of Chicago, Chicago, IL, 60603, USA}
\altaffiltext{\CfA}{Harvard-Smithsonian Center for Astrophysics, Cambridge, MA, 02138, USA}
\altaffiltext{\DI}{Dunlap Institute for Astronomy and Astrophysics, University of Toronto, 50 St George St, Toronto, ON, M5S 3H4, Canada}
\altaffiltext{\Toronto}{Department of Astronomy and Astrophysics, University of Toronto, 50 St George St, Toronto, ON,  M5S 3H4, Canada}
\altaffiltext{\BCCP}{Berkeley Center for Cosmological Physics, Department of Physics, University of California, Berkeley, CA, 94720, USA}

\label{firstpage}

\begin{abstract}We measure the cross-power spectrum of the projected mass
density as traced by the convergence of the cosmic microwave background
lensing field from the South Pole Telescope (SPT) and a sample of Type 1 and 2
(unobscured and obscured) quasars at $\langle z \rangle \sim 1$ selected with
the {\it Wide-field Infrared Survey Explorer}, over 2500\,deg$^2$. The
cross-power spectrum is detected at $\approx$7$\sigma$, and we measure a
linear bias $b=1.61\pm0.22$, consistent with clustering analyses. Using an
independent lensing map, derived from {\it Planck} observations, to measure
the cross-spectrum, we find excellent agreement with the SPT analysis. The
bias of the combined sample of Type 1 and 2 quasars determined in this work is
similar to that previously determined for Type 1 quasars alone; we conclude
that obscured and unobscured quasars trace the matter field in a similar way.
This result has implications for our understanding of quasar unification and
evolution schemes. \end{abstract} \keywords{cosmology: observations}

\section{Introduction}

The trajectories of photons that comprise the cosmic microwave background
(CMB) have been gravitationally deflected by large scale structure. The
observational consequence is a smoothing of the CMB temperature power
spectrum, and the introduction of correlations between what were originally
independent modes. These effects allow one to map the total projected
gravitational potential of the universe back to the surface of last
scattering.

CMB experiments have now achieved the sensitivity and resolution to directly
detect the non-Gaussian signature left on the CMB by gravitational lensing
(Das et al.\ 2011; van\ Engelen et al.\ 2012, Planck Collaboration\ 2013\
XVII). A method of mapping the lensing potential is the optimal quadratic
estimator (Seljak \& Zaldarriaga 1999; Hu\ 2001) that allows one to separate
the lensing perturbation from the intrinsic power spectrum. The redshift where
the weight of the lensing kernel peaks is close to the maximum in the global
volume-averaged star formation and black hole growth rates; CMB lensing
studies therefore promise exciting new insights into the complex relationship
between the growth of luminous galaxies and the dark matter overdensities they
inhabit, in additional to cosmological applications (Smith et al.\ 2007;
Hirata et al.\ 2008; Sherwin et al.\ 2012; Bleem et al.\ 2012; Holder et al.\
2013; Planck Collaboration\ 2013\ XVIII).

Quasars are visible over cosmological distances even in relatively shallow
surveys, and have a long history as cosmological probes. They also represent
an important phase in the evolution of massive galaxies, since their
luminosities arise from an episode of supermassive black hole growth. How this
phase dovetails with the global scheme of galaxy evolution remains to be fully
understood. Sherwin et al.\ (2012) presented the first detection of a
significant (3.8$\sigma$) cross-correlation signal between Sloan Digital Sky
Survey (SDSS) optically-selected quasars and the CMB lensing convergence
measured by the Atacama Cosmology Telescope (ACT) over 320\,deg$^2$. Here we
use the {\it Wide-field Infrared Survey Explorer} ({\it WISE}; Wright et al.\
2010) to select both Type 1 and Type 2 (unobscured and obscured) quasars over
a 2500\,deg$^2$ field for which we have a map of the CMB measured by the South
Pole Telescope (SPT; Carlstrom et al.\ 2011; Story et al.\ 2012).

In this Letter, we present a cross-correlation analysis examining the
relationship between the combined Type 1 and 2 quasar population and the
matter field. A $\Lambda$CDM cosmology defined by the parameters measured with
the {\it Wilkinson Microwave Anisotropy Probe} (7 year results including
baryonic acoustic oscillation and Hubble constant constraints; Komatsu et al.\
2011) is assumed and throughout {\it WISE} fluxes are quoted on the Vega
magnitude system.

\section{Data}

\subsection{SPT lensing convergence map}

The SPT temperature survey (Carlstrom et al.\ 2011) covers $2500$\,deg$^2$
($\alpha=20^{\rm hr}\rightarrow7^{\rm hr}$, $\delta=-65^\circ\rightarrow
-40^\circ$) at $\nu_{\rm obs}=95, 150$, and $220$\,GHz to typical 1$\sigma$
depths of 40, 18, and 70 $\mu$K-arcmin. For this work we only use 150\,GHz
data. The full survey comprises many individual fields which are combined for
this work to make CMB temperature maps that are 17\,degrees on a side. Maps of
the gravitational potential (van Engelen et al.\ 2012) for these 17 degree
fields are mosaiced into a single 2500\,deg$^2$ map at 3$'$ resolution. We use
a single lensing filter for all regions of the SPT survey; because the noise
fluctuates mildly from field to field, this is not optimal. For precise
measures of the lensing power spectrum (van Engelen et al. 2012; Zahn et al in
prep) optimal lensing filters are made using the individual noise levels of
each of the original 100\,deg$^2$ individual fields.

This procedure is repeated for 40 sets of simulations, which consist of lensed
CMB maps and realistic SPT noise, following the procedure outlined in van
Engelen et al.\ (2012). The resulting full-field simulated gravitational
lensing maps are cross-correlated with the input maps to obtain an effective
transfer function that can be used to correct the observed cross-correlations
(\S4.1). As part of this validation procedure, we also tested the reliability
of the `flat sky' approximation. Projecting the input full-sky gravitational
potential maps into our Zenithal Equal Area projection, we verified that the
mean power spectra of the projected 2500\,deg$^2$ maps agree with the full-sky
curved-sky power spectrum to much better than a few per cent for $l>20$, and
that power spectra of 2500\,deg$^2$ flat sky maps agreed to similar precision
with curved-sky power spectra.

Any correlation between the CMB foreground power and the galaxy density
can lead to a small bias in the CMB lensing-galaxy cross-power on large scales
(van Engelen et al.\ 2012; Bleem et al.\ 2012). For infrared intensity
fluctuations or the thermal Sunyaev--Zel'dovich effect, this was found to be
on the order of a few percent (van Engelen et al.\ 2012), and it is not
expected to be substantially larger for quasars.

\subsection{WISE quasar selection}

\subsubsection{Selection of quasars at 3.4--4.6$\mu$m}

{\it WISE} has mapped the sky at 3.4, 4.6, 12, and 22$\mu$m ({\it W1}--{\it
W4}), and offers a unique resource to study the demographics of quasars and
active galactic nuclei (AGN)\footnote{In this work we use the term `quasar' to
refer to both classical quasars and AGN since the majority of the objects in
our sample have high luminosities characteristic of `quasars'.}.

We create a catalog of galaxies from the {\it WISE} All-sky
Release\footnote{http://irsa.ipac.caltech.edu/Missions/wise.html}, selecting
all sources with (a) signal-to-noise in {\it W2} {\it w2snr}\,$\geq$\,10, (b)
{\it W2} magnitude {\it w2mpro}\,$\leq$\,15\,mag, (c) data quality flags {\it
cc\_flags}\,$=$\,`0' in both the {\it W1} and {\it W2} bands, (d) number of
PSF components fit {\it nb}\,$=$\,1, and (e) number of active deblends {\it
na}\,$=$\,0. The choice of signal-to-noise (a) and magnitude limit (b) in the
{\it W2} (4.6$\mu$m) band is to ensure approximately uniform completeness
across the full SPT footprint. The data quality parameter (c) ensures that
contamination of the catalog by false detections and errors in PSF-fit
photometry from instrumental artifacts is minimized. We also require that only
a single PSF component is fit to measure the photometry (d) and the source has
not been actively de-blended (e), improving reliability.

The selection of quasars in the mid-infrared is a well-established technique
(Lacy et al.\ 2004, Stern et al.\ 2005). Assef et al.\ (2012) show that in the
{\it WISE} bands a selection of $W1-W2\geq0.8$ and $W2\leq15$\,mag returns a
reliable sample of both Type 1 and 2 AGN. Figure\ 1 illustrates the efficacy
of this selection in the {\it W1-W2-W3} color plane, using spectroscopically
classified sources selected from the Sloan Digital Sky Survey (Data Release 7,
Abazajian et al.\ 2009), matched to {\it WISE}. With the {\it WISE} AGN cut,
we select 107,469 objects, corresponding to a surface density of
42\,deg$^{-2}$.

\begin{figure} \includegraphics[height=0.5\textwidth,angle=-90]{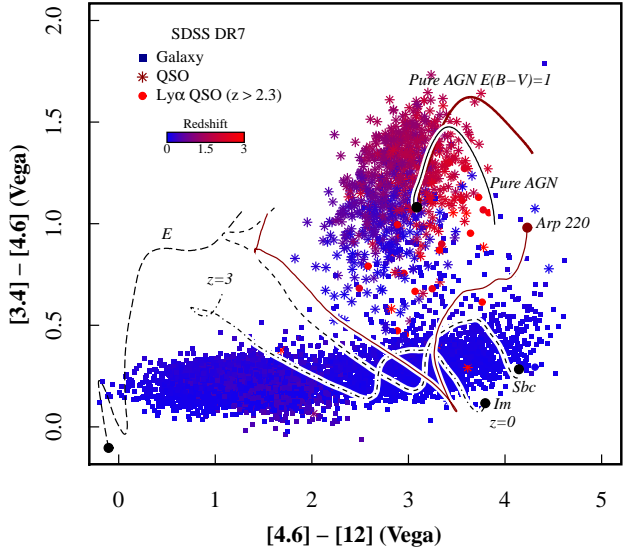}
\caption{{\it WISE} {\it W1-W2-W3} (3.4-4.6-12$\mu$m) colors of 6000
spectroscopically classified galaxies and quasars from SDSS. Tracks show the
colors of typical normal galaxies (elliptical [E], mid-type spiral [Sbc] and
irregular/starburst [Im]), a prototypical ultraluminous starburst galaxy (Arp
220) and a zero star formation quasar seen at $0<z<3$ (Polletta et al.\ 2007;
Vega et al.\ 2008; Assef et al.\ 2013; Yan et al.\ 2013). We show the effect
of internal reddening on the quasar track with an extinction of
$E(B-V)=1$\,mag. This demonstrates how {\it WISE} colors can be used to
cleanly separate quasars from normal galaxies.} \end{figure}

\subsubsection{Redshift distribution}

To obtain an estimate of the redshift distribution of our sample, we select
objects from the {\it WISE} All-sky Survey catalog using the criteria
described above within the 9 deg$^2$ Bo\"otes survey field, which has
extensive spectroscopy as part of the 7.9 deg$^2$ AGN and Galaxy Evolution
Survey (AGES; Kochanek et al.\ 2012) and photometric redshift estimates using
optical and {\it Spitzer} IRAC imaging (Brodwin et al.\ 2006).

Figure\ 2 shows the redshift distribution of 379 {\it WISE} sources matching
our selection in the AGES survey region with spectroscopic redshifts (89\%) or
photometric redshift estimates ($\delta z\approx 0.3$). The distribution peaks
at $\langle z \rangle = 1.1$ and has a spread of $\Delta z = 0.6$. We have
redshift estimates for 93\% of the {\it WISE}-selected AGN in the AGES region,
and so this is likely to be a robust model of the redshift distribution of
identically-selected sources in the SPT field. Note that the
spectroscopic Bo\"otes data allows us to estimate the contamination rate from
non-AGN in our selection, which is $\lesssim$15\%.

\begin{figure} \includegraphics[height=0.5\textwidth,angle=-90]{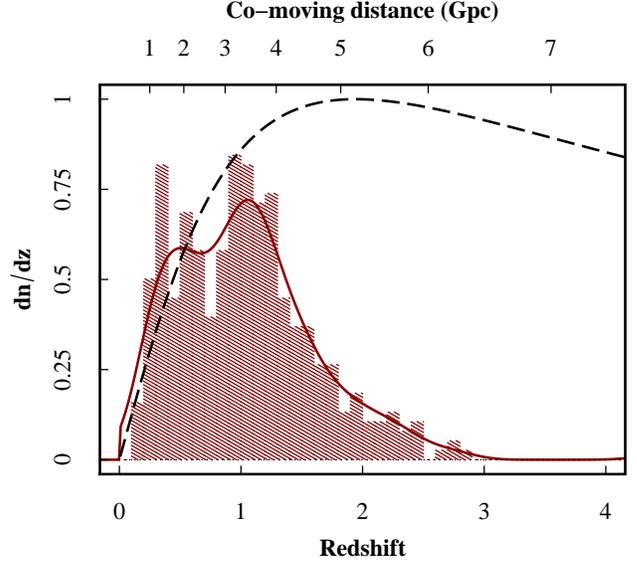}
\caption{The normalized redshift distribution of 379 quasars selected using
the criteria $W1-W2\geq0.8$ and $W2\leq15$\,mag in the 7.9\,degree$^2$ Bo\"otes/AGES
field (\S2.2.2). The solid line shows the density estimate of the discrete
redshift distribution which we use as the model $dn/dz$. The dashed line shows
the shape of the CMB lensing kernel, plotted as $(d\chi/dz)W^\kappa$ (\S3).}
\end{figure}

\begin{figure*} \centerline{\includegraphics[width=\linewidth]{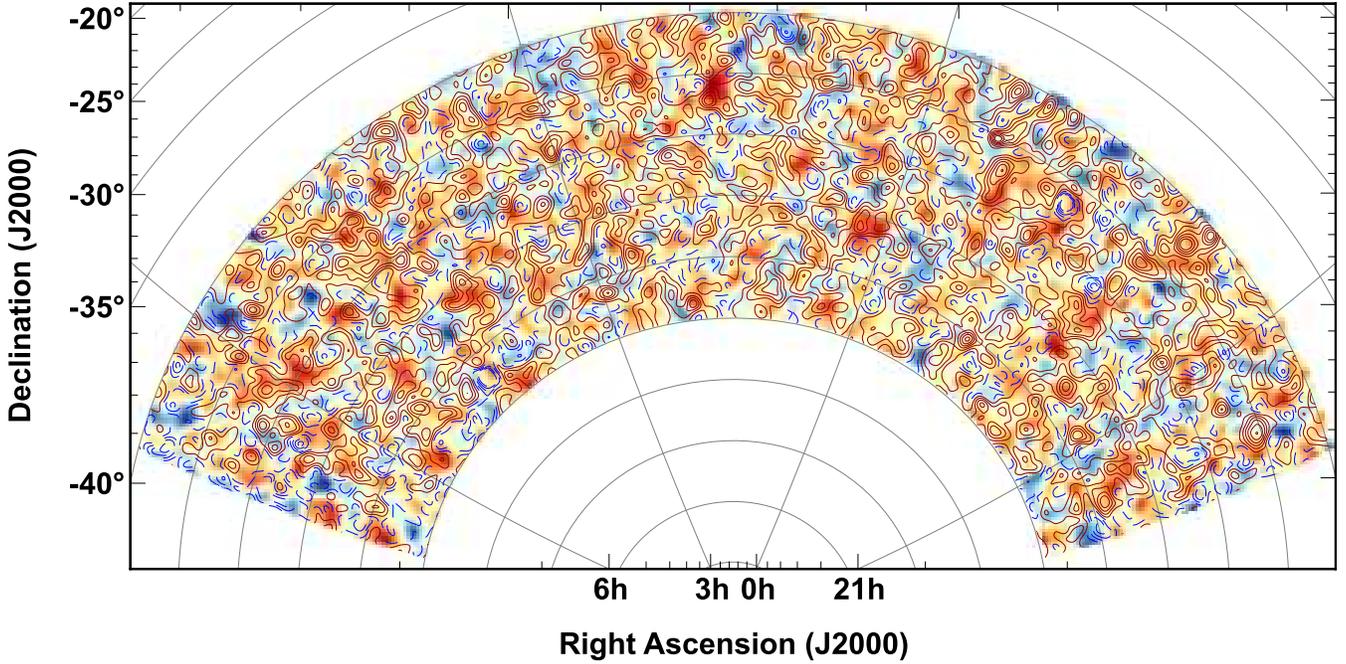}}
\caption{The 2500\,deg$^2$ SPT lensing convergence map with contours showing
the fractional overdensity of quasars, both smoothed with a 1$^\circ$ Gaussian
kernel. The color scale runs from blue$\rightarrow$red for regions with
negative to positive relative convergence (see \S3). Contours span
$-0.5\leq\delta\leq 0.5$ in steps of $\delta=0.1$; $\delta<0$ contours are
dashed. The CMB lensing convergence map and the quasar number density field
are correlated at the 7$\sigma$ level.} \end{figure*}

\section{CMB lensing theory review}

We briefly review the formalism presented in Bleem et al.\ (2012). The lensing
convergence $\kappa=-\nabla\cdot\mathbf{d}/2$ (where $\mathbf{d}$ is the
deflection field) along a line of sight $\hat{\mathbf n}$, 
can be expressed as the integral, over comoving distance $\chi$, 
of the fractional overdensity
of matter $\delta(\mathbf{r},z)$, multiplied by the lensing kernel, $W^\kappa$:

\begin{equation} \kappa(\hat{\mathbf n}) = \int d\chi
W^\kappa(\chi)\delta(\chi{\hat{\mathbf n}},z(\chi)), \end{equation}

\noindent where the lensing kernel is (Cooray \& Hu\ 2000; Song et al.\ 2003):

\begin{equation} W^\kappa(\chi) = \frac{3}{2}\Omega_{\rm
m}H_0^2\frac{\chi}{a(\chi)}\frac{\chi_{\rm CMB} - \chi}{\chi_{\rm CMB}}.
\end{equation} 

\noindent Here $\Omega_{\rm m}$ and $H_0$ are the present-day values of the
ratio of the matter density to the critical density and Hubble parameter
respectively, and $a(\chi)$ is the scale factor. The comoving distance to the
surface of last scattering $\chi_{\rm CMB}\approx14$\,Gpc in our cosmology.

Quasars, like all galaxies, are biased tracers of the matter field, and so
fluctuations in the galaxy density can be expressed

\begin{equation} g(\hat{\mathbf n}) = \int d\chi
W^g(\chi)\delta(\chi{\hat{\mathbf n}},z(\chi)) \end{equation}

\noindent where $W^g(\chi)$ is the quasar distribution kernel

\begin{equation} W^g(\chi) = \frac{dz}{d\chi}\frac{dn(z)}{dz} b(\chi)
\end{equation}

\noindent where $dn(z)/dz$ is the normalized redshift distribution of the
population, and $b$ is the bias. Comparing equations 2 \& 4, the lensing
quantity analogous to $dn/dz$ is $(d\chi/dz)W^\kappa$. This is plotted in
Figure 2 for comparison. With these defined, the cross-power at angular 
frequency $l$, assuming the Limber approximation 
(Limber\ 1953, Kaiser\ 1992), is

\begin{equation} C^{\kappa g}_l = \int dz
\frac{d\chi}{dz}\frac{1}{\chi^2}W^\kappa(\chi)W^g(\chi)P\left(\frac{l}{\chi},z
\right) \end{equation}

\noindent where $P(k=l/\chi,z)$ is the non-linear matter power spectrum, which
we generate from the Code for Anisotropies in the Microwave Background (CAMB;
Lewis et al.\ 2000, online
version\footnote{http://lambda.gsfc.nasa.gov/toolbox/tb\_camb\_form.cfm}) which
calculates the non-linear matter power spectrum using HALOFIT (Smith et al.\
2003).

\section{Analysis and results}

A quasar density map, expressed as $\delta = (\rho-\langle\rho\rangle)/\langle
\rho \rangle$, is generated from the {\it WISE} catalog on a grid matching the
3$'$/pixel scale of the convergence map (Figure\ 3). Both the quasar map and
the SPT lensing map have been smoothed to show scales where the lensing
convergence map has significant signal-to-noise.

If quasars are tracing peaks in the matter density field, then we would expect
that on average the convergence will be enhanced in regions of high quasar
density (and will be lower in regions of low density). This should be apparent
in a `stack' of the convergence map at different positions of the density map.
We define 10 density bins covering the range $-0.5\leq \delta \leq 0.5$, split
such that each bin represents the same sky area. The average $\kappa$ for each
bin is then

\begin{equation}\bar{\kappa}|_\delta=\frac{1}{N}\sum_{i=0}^{N} \kappa(x_i,y_i)\end{equation}

\noindent where $\kappa(x_i,y_i)$ is the value of $\kappa$ at the $i$th pixel
in each $\delta$ bin. The significance of the stack can be estimated by
repeating the procedure on an ensemble of 40 realistic noise simulations and
taking the variance for each bin. Figure\ 4 presents the stacked images,
showing a significant transition from mean negative convergence in regions of
low quasar density to positive convergence in regions of high quasar density.
This is clear evidence that {\it WISE}-selected quasars are tracing mass that
is lensing the CMB.

\begin{figure*}[t] \centerline{\includegraphics[width=\textwidth]{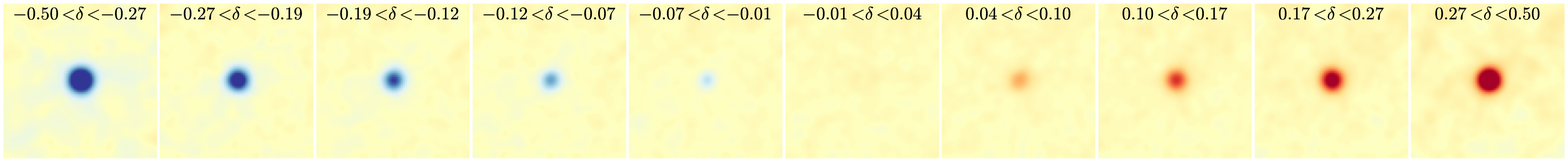}}
\centerline{\includegraphics[width=\textwidth]{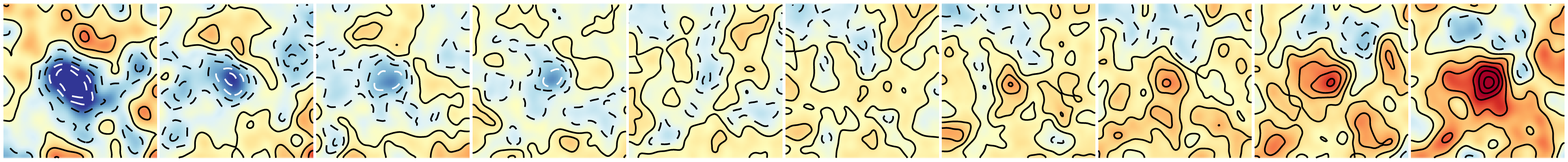}} \caption{$5^\circ$
thumbnail stacks showing (top) bins of fractional quasar density where we have
stacked the $\delta$ map in bins spanning $-0.5\leq \delta \leq 0.5$; (bottom)
equivalent stacks evaluated at the same positions in the lensing convergence
map. Contours show the significance in levels of 1$\sigma$, based on
simulations (\S2). Dashed contours indicate the significance in regions of
$\kappa<0$. A clear, significant transition from negative to positive CMB
lensing convergence for lines of sight to low$\rightarrow$high relative quasar
density is evident, graphically illustrating the strong cross-correlation
signal.} \end{figure*}

\subsection{Cross-power spectrum}

The cross-correlation is comprehensively measured through the cross-power
spectrum (equation 5). Since the redshift distribution of the quasar
population is reasonably well constrained (\S2.2.2), this allows us to
estimate the bias of the population. In Figure\ 5 we present the cross-power
spectrum of the convergence map, ${\sf M_\kappa}$, and quasar density map
${\sf M_g}$:

\begin{equation} C_l^{\kappa g} = \left< \mathrm{Re}( \mathcal{F}({\sf
M_\kappa})\mathcal{F^*}({\sf M_g}))|_{\mathbf{l}\in l} \right>
\end{equation}

\noindent where $\mathbf{l}\in l$ describes the binning, such that the average
power is calculated over all pixels in 2d Fourier space with coordinate
$\mathbf{l}$ within the bin defined by $l$. As in Bleem et al.\ (2012), we
mask bright stars identified by 2MASS. When evaluating $C_l^{\kappa g}$ in
bins of $l$, we correct for the transfer function described in \S2.1,
corresponding to a factor of $\approx$10--30\% for the bins shown.

Uncertainties are derived by repeating the calculation with 40 realistic noise
simulations (\S2.1) in place of the real convergence map and measuring the
variance in $C_l^{\kappa g}$, where the mean correlation between the
simulations and the quasar map is null for all $l$. Note that this
procedure underestimates the sample variance contribution by a factor of
$\surd 2$, but since the quasar catalogue is shot-noise dominated, neglecting
this will underestimate the uncertainty by less than 10\%. Fitting equation
(5) for the bias, we find a best fit $b=1.61\pm0.22$, with $\chi^2=1.32$ and
$\chi^2/\nu=0.26$. The significance is evaluated as the difference between the
null line ($b=0$) and the best fit theoretical spectrum: $\Delta \chi^2 =
\chi^2_{\rm null}-\chi^2_{\rm fit}$, corresponding to a detection significance
of $7.0\sigma$. It is possible to obtain a more significant (13$\sigma$)
signal if one simply cross-correlates a `generic extragalactic' sample defined
by $15\leq W1\leq 17$\,mag (e.g.,\ Bleem et al.\ 2012), however, this
population is so heterogeneous that it is difficult to interpret any derived
parameters for the galaxies involved.

\subsubsection{Planck comparison}

The availability of {\it Planck} data allows us to follow an identical
procedure using an independent lensing map. An all-sky lensing potential map
from {\it Planck} (Planck Collaboration\ 2013 XVII) retrieved from the {\it
Planck} Legacy Archive is converted to lensing convergence using spherical
harmonic transforms, then projected onto the SPT survey area. Figure\ 5 shows
the excellent agreement between SPT and {\it Planck} cross-power spectra.
Without realistic {\it Planck} simulations it is difficult to accurately
estimate uncertainties, so the {\it Planck} error bars are derived from the
$l$ bin variance, which we have verified (using the SPT power spectrum) is a
good estimate of the uncertainty derived from noise simulations. 
Generally, uncertainties in the {\it Planck} spectrum are $\sim$20\% larger
than SPT, however note that shot-noise in the quasar catalog is a significant
contribution in the errors of both spectra. Furthermore, the strongly
anisotropic noise in the lensing map (van Engelen et al.\ 2012) is not
included in the cross-spectrum estimation, leading to sub-optimal power
spectrum estimates, and therefore more similar error bars for {\it Planck} and
SPT than might be naively expected.

\section{Interpretation} 

In $\Lambda$CDM, one can relate galaxy populations to dark matter halos of
characteristic mass $M_{\rm h}$ (e.g.,\ Peebles\ 1993), in the simplified case
in which all objects in a given sample reside in halos of the same mass.
$M_{\rm h}$ is related to the bias through the parameterization $b = f(\nu)$
where $\nu$ is the ratio of the critical threshold for spherical collapse to
the r.m.s. density fluctuation for a mass $M$: $\nu =\delta_c/\sigma(M)$. 
Here we apply the fitting function of Tinker et al.\ (2010)\footnote{assuming
halos are all 200 times the mean density of the universe}, yielding
$\log_{10}(M_{\rm h}/[h^{-1}M_\odot])=12.3^{+0.3}_{-0.2}$ for our measured
$b=1.61\pm0.22$, at $\langle z\rangle=1.1$.

Sherwin et al.\ (2012) used galaxy-CMB lensing cross-correlation to measure
the linear bias of SDSS photometrically-selected Type 1 quasars, finding $b =
2.5 \pm 0.6$ at $z \approx 1.5$. The bias of Type 1 quasars is observed to
evolve over our redshift range (Croom et al.\ 2005; Ross et al.\ 2009), and
indeed Sherwin et al.\ (2012) assume a fiducial model for $b(z)$ in their fit.
The evolution of the bias of Type 1 and 2 quasars {\it combined} is not known;
but if we assume a fiducial evolution model for $b(z)$ appropriate for Type 1,
$b_{\rm evo}=0.53 + 0.289(1+z)^2$ (Croom et al.\ 2005) and fit for the
normalization of that model, $b = b_{\rm 0}b_{\rm evo}$, we find $b_{\rm
0}=0.97\pm0.13$, corresponding to $b=1.75\pm 0.23$ at $\langle z \rangle=1.1$
($\log(M_{\rm h}/[h^{-1}M_\odot]) = 12.4^{+0.2}_{-0.3}$) (Fig.\ 5). This is in
excellent agreement with that of clustering analyses of Type 1 quasars
($b=1.83\pm0.33$ at $z\approx1$, Ross et al.\ 2009). Evolving this model for
bias evolution to $z\approx1.5$, we obtain $b=2.27$, consistent with Sherwin
et al.\ (2012).

The key difference between previous studies and ours is the fact that the
{\it WISE} selection includes both Type 1 and Type 2 quasars. Due to the
current paucity of deep optical data across the SPT footprint, we are unable
to split our sample into Type 1 and 2 (Hickox et al.\ 2007). However, using
our identical Bo\"otes selection (\S2.2.2), where a Type classification {\it
can} be made (which is generally at $z>0.7$), we find similar $dn/dz$, with
$\left<z\right>=1.21$ and $\left<z\right>=1.11$ for Type 1 and 2 quasars (with
similar mean bolometric luminosities of $\log(L_{\rm bol}/{\rm
erg\,s^{-1}})=46.18$ and 46.16) respectively (Hickox et al.\ 2011). Assuming
the similarity in the redshift distribution of Type 1 and 2 quasars persists
to $z<0.7$, then our result implies that Type 1 and 2 quasars trace the matter
field in a similar way, given the similarity with the bias measured for Type 1
quasars alone.

The relative abundance of Type 1 and 2 quasars in our selection is
$\sim$70:30, however, intrinsically they are thought to be approximately
equally abundant (Ueda et al.\ 2003, Hopkins et al.\ 2007). This is explained
through our bright cut in $W2$ (and strict $W1-W2$ selection), introducing
incompleteness that preferentially affects the obscured quasars. The
conclusions that follow make the assumptions that (a) Type 1 and 2 quasars are
equally abundant and have similar redshift distributions, and (b) our bias
measurement is representative of the population as a whole, at the bolometric
luminosities sampled here.

\begin{figure}[t]
\centerline{\includegraphics[height=0.5\textwidth,angle=-90]{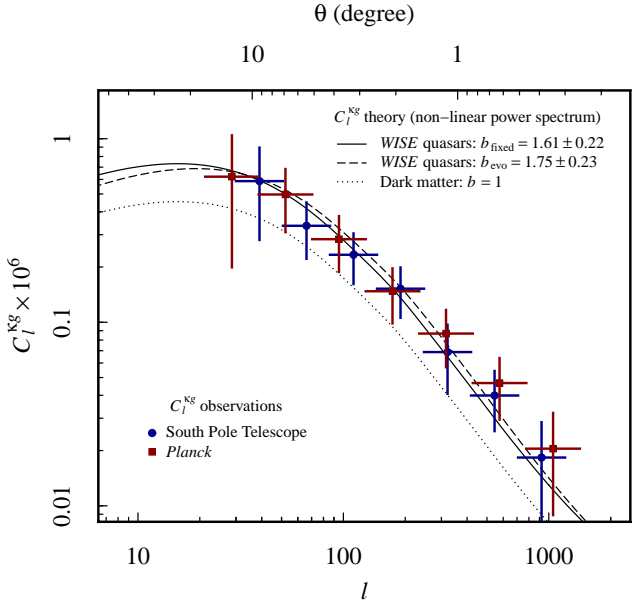}}
\caption{Cross-power spectrum of the {\it WISE}-selected quasar density and
the CMB lensing convergence. The curves show (i) dark matter ($b=1$, dotted),
(ii) the best-fit (to SPT) equation 5, (solid), with constant bias and (iii)
evolving bias (dashed, \S5).} \end{figure}

\section{Conclusions}

Our result shows that the bias of a combined Type 1 and 2 quasar sample is
consistent with that found for a Type 1 sample alone at $z\sim1$. This is in
agreement with Hickox et al.\ (2011), who conclude that Type 2 quasars must be
at least as strongly clustered as Type 1 quasars. This is important for quasar
evolution and unification schemes. In the unification model, Type 1 and 2
quasars are fundamentally the same population, but the geometry of the
material obscuring the optically bright accretion disc results in an optical
depth that is strongly dependent on viewing angle. In unification we would
expect to find that the bias of a mixture of Type 1 and Type 2 quasars is the
same as a Type 1-only sample selected from the same redshift distribution.

An alternative hypothesis is that obscured quasars become unobscured through a
process that removes the optically thick nuclear screen (e.g.,\ Hopkins et
al.\ 2008). The relative bias of the populations can be related to the physics
of the transitionary process. The similar bias parameters imply similar masses
for the host halos, and therefore a comparable host halo number density. If
the intrinsic abundances of obscured and unobscured quasars are roughly equal,
and their bolometric luminosities similar, then this implies that the obscured
and unobscured evolutionary phases must be of similar duration.

The technique of galaxy--CMB lensing cross-correlation is an exciting and
powerful new tool for examining the complex relationship between luminous
galaxies and the dark matter field they inhabit.

\medskip

\noindent SPT is supported by the NSF through grants ANT-0638937 and
ANT-0130612. Support for this work is provided by: the NSF Physics Frontier
Center grant PHY-0114422 to the Kavli Institute of Cosmological Physics at the
University of Chicago; the Kavli Foundation and the Gordon and Betty Moore
Foundation; NSF (grant numbers 1211096, 1211112 and PHYS-1066293); NASA
through ADAP award NNX12AE38G; NSERC, CIfAR, and the Canada Research Chairs
program. This research used resources of the National Energy Research
Scientific Computing Center, which is supported by the Office of Science of
the U.S. DoE under contract DE-AC02-05CH11231. Research at Argonne National
Laboratory is supported by the Office of Science of the U.S. DoE under
contract DE-AC02-06CH11357. {\it Planck} is an ESA science mission with
instruments and contributions directly funded by ESA Member States, NASA, and
Canada.

\label{lastpage} 
\begin{thebibliography}{100}

\bibitem[Author et al. (2012)]{author}{Abazajian, K., et al.,\ 2009, ApJS,
182, 543}

\bibitem[Author et al. (2012)]{author}{Assef, R. J., et al., 2012, ApJ submitted
arXiv:1209.6055}

\bibitem[Author et al. (2012)]{author}{Bleem, L. E., et al., 2012, ApJ, 753,
L9}

\bibitem[Author et al. (2012)]{author}{Brodwin, M. J. I., et al. 2006, ApJ, 651, 791}

\bibitem[Author et al. (2012)]{author}{Carlstrom, J. E.., et al., 2011, PASP, 123, 568}

\bibitem[Author et al. (2012)]{author}{Croom, C. M., Boyle, B. J.; Shanks, T.; Smith, R. J., Miller, L., Outram, P. J., Loaring, N. S., Hoyle, F., da \^Angela, J. 2005, MNRAS, 356, 415}

\bibitem[Author et al. (2012)]{author}{Cooray, A., Hu, W., 2000, ApJ, 534, 533}

\bibitem[Author et al. (2012)]{author}{Das, S., et al.,\ 2011, PRL, 107, 2, 021301}

\bibitem[Author et al. (2012)]{author}{Eisenstein, D. J., Hu, W., 1999, ApJ,
511, 5}

\bibitem[Author et al. (2012)]{author}{Hickox, R. C., et al., 2007, ApJ, 671, 1365}

\bibitem[Author et al. (2012)]{author}{Hickox, R. C., et al., 2011, ApJ, 731, 117}


\bibitem[Author et al. (2012)]{author}{Hirata, C. M., Ho, S., Padmanabhan, N., Seljak, U., Bahcall, N. A., 2008, PRD, 78, 4, 043520}

\bibitem[Author et al. (2012)]{author}{Holder, G. P., et al., 2013, ApJL in
press, 2013arXiv1303.5048}

\bibitem[Author et al. (2012)]{author}{Hopkins, P. F.; Richards, G. T., Hernquist, L., 2007, ApJ, 654, 731}
	
\bibitem[Author et al. (2012)]{author}{Hopkins, P. F., Hernquist, L., Cox, T. J., Keres, D., 2008, ApJS, 175, 356}

\bibitem[Author et al. (2012)]{author}{Hu, W., 2001, ApJ, 557, L79}

\bibitem[Author et al. (2012)]{author}{Kaiser, N., 1992, ApJ, 388, 272}

\bibitem[Author et al. (2012)]{author}{Kochanek, C. S. et al. 2012, ApJS, 200,
8}

\bibitem[Author et al. (2012)]{author}{Komatsu, E., et a., 2011, ApJS, 192, 18}

\bibitem[Author et al. (2012)]{author}{Lacy, M., et al., 2004, ApJS, 154, 166}

\bibitem[Author et al. (2012)]{author}{Lewis, A., Challinor, A.,  Lasenby, A. 2000, ApJ, 538, 473}

\bibitem[Author et al. (2012)]{author}{Limber, D. N, 1953, ApJ, 117, 134}



\bibitem[Author et al. (2000)]{author}{Peebles, P. J. E., 1993, Principles of physical cosmology, Princeton University Press, Princeton, NJ}

\bibitem[Author et al. (2013)]{author}{Planck Collaboration XVII. 2013, 
arxiv:1303.5077}

\bibitem[Author et al. (2013)]{author}{Planck Collaboration XVIII. 2013, 
arxiv:1303.5078}

\bibitem[Author et al. (2012)]{author}{Polletta, M., et al., 2007, ApJ, 663,
81}


\bibitem[Author et al. (2012)]{author}{Ross, N. P., et al., 2009, ApJ, 697,
1634}

\bibitem[Author et al. (2012)]{author}{Seljak, Uros, Zaldarriaga, Matias, 1999, PRL, 82, 13, 2636}


\bibitem[Author et al. (2012)]{author}{Sherwin, B. D., et al., 2012, PRD, 86, 8, 083006}

\bibitem[Author et al. (2012)]{author}{Song, Y.-S., Cooray, A., Knox, L.,
Zaldarriaga, M., 2003, ApJ, 590, 664}

\bibitem[Author et al. (2012)]{author}{Smith, R. E., Peacock, J. A., Jenkins, A., et al. 2003, MNRAS, 341, 1311}

\bibitem[Author et al. (2012)]{author}{Smith, K. M.; Zahn, O.; Dor\'e, O., 2007, PRD, 76, 4, 043510}


\bibitem[Author et al. (2012)]{author}{Stern, D., et al., 2005, ApJ, 631, 163}

\bibitem[Author et al. (2012)]{author}{Story, K. T., et al., arXiv1210.7231}

\bibitem[Author et al. (2000)]{author}{Tinker, J. L., Robertson, B. E., Kravtsov, A. V., Klypin, A., Warren, A. S., Yepes, G., Gottl\"ober, S., 2010 ApJ, 724, 878}

\bibitem[Author et al. (2000)]{author}{Ueda, Y., Akiyama, M., Ohta, K., Miyaji, T., 2003, ApJ, 598, 886}
        
\bibitem[Author et al. (2012)]{author}{van Engelen, A., et al.\ 2012, ApJ, 756, 142}

\bibitem[Author et al. (2012)]{author}{Vega, O., Clemens, M. S., Bressan, A., Granato, G. L., Silva, L., Panuzzo, P., 2008, A\&A, 484, 631 }


\bibitem[Author et al. (2012)]{author}{Wright, E. L., et al., 2010, AJ, 140, 1868}

\bibitem[Author et al. (2012)]{author}{Yan, L., et al., 2013, AJ, 145, 55}

\end{thebibliography}
\end{document}